\documentstyle[12pt,aps]{revtex}
\newcommand{\be}{\begin{equation}}
\newcommand{\ee}{\end{equation}}
\newcommand{\ba}{\begin{eqnarray}}
\newcommand{\ea}{\end{eqnarray}}
\newcommand{\oh}{\displaystyle{\frac{1}{2}}}
\begin{document}
\title{Fixed Charge Ensembles and Induced Parity Breaking Terms.}
\author{C.D.~Fosco$^a$\thanks{CONICET}
\\
{\normalsize\it
$^a$Centro At\'omico Bariloche,
8400 Bariloche, Argentina}}
\maketitle
\begin{abstract}
Recently derived results for the exact induced parity-breaking term 
in 2+1 dimensions at finite temperature are shown to be relevant to the 
determination of the free energy for fixed-charge ensembles. The partition functions for fixed 
total charge corresponding to massive fermions in the presence of Abelian
and non-Abelian magnetic fields are discussed. We show that the presence 
of the induced Chern-Simons term manifests itself in that the free energy 
depends strongly on the relation between the external magnetic flux and 
the value of the fixed charge.
\end{abstract}
\begin{center}
\end{center}
\bigskip
\newpage
Quantum Field Theory in $2+1$ dimensions continues to be a subject 
of active research, because of its many distinctive properties,
with no $3+1$ dimensional counterpart. Examples, provided by 
$2+1$-dimensional (`planar') models, are worth studying not only by 
purely theoretical reasons, but also because many important physical 
systems, or experimental situations, are indeed essentially 
planar, as in the well-known examples borrowed from Condensed Matter 
Physics~\cite{frad}. This also happens for some astrophysical objects whose 
configurations are approximately invariant under translations along 
one of the spatial dimensions, what renders the relevant dynamics 
two-dimensional. 
One of the more striking properties of $2+1$ dimensional physics is 
that it allows for the existence of fractional statistics, realized 
in terms of the so-called `anyons', namely, identical particles with 
neither bosonic nor fermionic statistics.
Closely related to fractional statistics is the fact that in $2+1$ 
dimensions a gauge field can be equipped with a gauge invariant 
and parity-breaking action with non-trivial topological properties, 
namely the Chern-Simons action. This action, if not introduced 
{\em ab initio\/} in the model, may be induced dynamically by 
virtual matter-field processes~\cite{djt}. The issue of the precise 
form of this induced action at finite temperature has been a 
long-standing problem, with obvious relevance for the applications. 
In some recent works, the exact expression for this induced term 
under some simplifying assumptions was derived for both the 
Abelian~\cite{dgs,frs1} and the non-Abelian cases~\cite{frs2}. 
This result was also rederived and generalized in \cite{af}.  
In this letter we shall show first that the configurations that 
have been studied in those references are precisely the ones needed 
in order to study the statistical mechanics of a fermion gas in a 
background magnetic field in the `fixed charge ensemble'~\cite{kapu}, 
and then we will find the difference between the free energy for 
such an ensemble and the one corresponding to the canonical ensemble. 
We will also show that properties like the behaviour of this 
induced action under large gauge transformations find a natural 
and concrete realization here. 
By a fixed-charge ensemble we shall  mean one where the values of 
one or more conserved and compatible (i.e., mutually commuting) charges 
have a delta-like statistical weight. Namely, if the fixed value is, 
say, $q$, only configurations having that eigenvalue for the charge 
operator are summed up in the statistical average. This should be 
contrasted with the grand-canonical ensemble, where only the 
{\em average\/} of the charge is fixed, but there is indeed room 
for fluctuations around this mean value. We may illustrate this 
distinction by saying that the microcanonical ensemble is a particular 
case of a fixed-charge one, where the fixed charge is just the 
Hamiltonian.
The interest in this kind of ensemble stems from the fact that the 
experimental situation under study may very well correspond to it 
(like in an electrically insulated sample, for example). The 
predictions shall differ significantly from the ones of other
ensembles for non-macroscopic systems 
(results will of course agree in the thermodynamic limit, where all 
the fluctuations may be ignored).  Illustrative examples of this kind 
of calculation are the colour singlet calculation 
(for an $SU(N)$ theory) of ref.~\cite{redl}, and
the fixed three-momentum ensemble of ref.~\cite{kapu}.

Our main idea in this letter is that, as the Chern-Simons term 
provides a link between the  magnetic field and the charge, it 
will strongly affect the statistical properties of a system in 
the presence of an external magnetic field, and in the fixed-charge 
ensemble. Moreover, we shall show that it is crucial to use the 
exact induced Chern-Simons term rather than the perturbative one 
in the derivation of this free energy.

The partition function ${\cal Z}_q$ corresponding to the ensemble
with fixed charge $q$, at a given temperature $T = \frac{1}{\beta}$,
for a system described by a quantum Hamiltonian $H$, and having a 
conserved additive charge $Q$ ($[H,Q] = 0$), is  
\be
{\cal Z}_q \;=\; \int_{-\pi}^\pi \, \frac{d \theta}{2 \pi} \,
                e^{-i \theta q} {\cal Z}_\theta
\label{defzq}
\ee
where
\be
{\cal Z}_\theta \;=\; {\rm Tr} e^{- \beta H + i \theta Q } \;.
\label{defzt}
\ee
We are assuming the normalization of $Q$ is such that its
eigenvalues are just integer numbers. Note that ${\cal Z}_\theta$
is formally equivalent to the grand canonical partition function
for a system with an {\em imaginary\/} chemical potential 
$\theta$. 
If the trace in (\ref{defzt}) is evaluated using a complete set
of simultaneous eigenstates of $H$ and $Q$, then it follows
immediately that (\ref{defzq}) will only pick up contributions
from quantum states with eigenvalue $q$ for $Q$. Also by using 
this complete set one sees that ${\cal Z}_\theta$ is a 
periodic function of $\theta$, with period $2 \pi$. Of 
course, this is closely related to the assumption that
particles in the physical spectrum have integer charge. We
shall see how this fact turns out to be important for the
application to the $2+1$ dimensional case, where this
periodicity is tantamount to gauge invariance under large
gauge transformations.
Alternatively, definition (\ref{defzq}) can be justified by
noting that 
\be
P_q \;=\; \int_{-\pi}^\pi \frac{d \theta}{2 \pi} 
          e^{-i \theta (q - Q)}           
\ee 
is a projector onto charge-$q$ states. In a fixed-charge ensemble,
the fixed charge does not fluctuate at all, as can be shown
explicitly by noting that the averages (denoted $\langle \cdots
\rangle_q$) of the powers of $Q$ may be written as
\be
\langle Q^n \rangle_q \;=\; (-i)^n {\cal Z}_q^{-1} \, 
\int_{-\pi}^\pi \frac{d \theta}{2 \pi} \, e^{-i \theta q} \,
\frac{\partial^n}{(\partial \theta)^n} {\cal Z}_\theta
\;=\; q^n \;,
\ee
where the periodicity of ${\cal Z}_\theta$ has been used in order
to ignore terms in the integration by parts. 
We want to construct the partition function ${\cal Z}_q (A)$ for the 
case of a fermionic field in $2+1$ dimensions in the presence of an
external magnetic field (here $A$ is the vector potential corresponding
to the magnetic field). From the analogy between 
${\cal Z}_\theta (A)$ and the partition function in the presence
of an imaginary chemical potential, we immediately obtain the
path-integral representation 
\be
{\cal Z}_\theta (A)\;=\; \int  {\cal D}{\bar \psi} {\cal D}\psi \,
\exp \left\{- \int_0^\beta d \tau \int d^2 x {\bar \psi} (\tau ,x)
[ \gamma_j D_j + M + \gamma_3 (\partial_\tau -i 
\frac{\theta}{\beta}) ] \psi (\tau ,x) \right\}
\label{zqa1}
\ee
where $D_j \;=\; \partial_j + i e A_j (x)$, and the notation and
conventions are identical to the ones used in \cite{frs1,frs2}.
It should now become evident that (\ref{zqa1}) corresponds to
exactly the same kind of configuration considered 
in \cite{frs1,frs2}, if one makes the identification 
${\tilde A}_3 = - \frac{\theta}{e \beta}$. 
Periodicity in $\theta$ for (\ref{zqa1}) is equivalent to invariance 
under large gauge transformations, after this identification is made.
We now separate ${\cal Z}_\theta$ into its phase and its modulus, 
which are given by the exponentials of the parity-breaking
and parity-conserving parts of the effective action,
respectively
\be
{\cal Z}_\theta \;=\; e^{-\Gamma_{odd}(A)} \times 
                      e^{-\Gamma_{even}(A)} \;.
\ee
We know from \cite{dgs,frs1,frs2} that, for this kind of configuration,
$\Gamma_{odd}$ can be exactly evaluated, and moreover that
its periodicity may be assured if the parity anomaly is
properly taken into account. As we have assumed that the 
ensemble corresponds to an {\em integer\/} charge $q$,  
periodicity of ${\cal Z}_\theta$ is required. We shall later on discuss 
the non-periodic `gauge anomalous' ${\cal Z}_\theta$.
The result for $\Gamma_{odd}$, including the parity anomaly 
piece is \cite{frs1,frs2}:
\be
\Gamma_{odd} (\theta, A) \;=\; i \frac{e}{2\pi} \frac{M}{|M|} \Phi 
\left\{ \; {\rm arctan} [\, \tanh(\frac{\beta |M|}{2})
     \tan (\frac{\theta}{2}) \,] \,-\, \oh \theta \;\right\} 
\label{godd}
\ee
where $\Phi \,=\,\int d^2 x \epsilon_{jk} \partial_j A_k $ 
is the static magnetic flux, and the branch of the ${\rm arctan}$
is chosen according to the value of $\theta$.
The even part of $\Gamma$ cannot be found exactly, but
fortunately there is a well-defined regime where its dependence
on $\theta$ can be safely ignored. This is the case when 
$\beta |M| >> 1$, as can be checked explicitly in
the calculation of \cite{af}, which yields the leading parity
conserving contribution to $\Gamma$. 
For example, in a smooth gauge field configuration (though the 
same holds true without this assumption),
\be
\Gamma_{even}(\theta,A_j) \, \simeq \, \Gamma^{(2)}(0,A_j)
\;+\; \frac{e^2 \beta}{48 \pi M}\;
\frac{{\rm tanh}(\frac{\beta M}{2})}{
{\rm cos}^2 (\frac{e \beta {\tilde A}_3}{2}) + 
{\rm tanh}^2 (\frac{\beta M}{2})
{\rm sin}^2 (\frac{e \beta {\tilde A}_3}{2})}
\int d^2 x F_{j k} F_{j k} \;, 
\label{qq}
\ee
where it becomes evident that dependence on ${\tilde A}_3$
(and hence on $\theta$) is exponentially suppressed for large 
$\beta |M|$. A more complete analysis shows that it is not even 
necessary to have $\beta |M| >>1$, but already for $\beta |M|$ of 
order $1$ the dependence on ${\tilde A}_3$ can be ignored.
Ignoring thus the $\theta$ dependence of $\Gamma_{even}$, 
\be
\Gamma_{even} (\theta, A_j) \;\simeq\; \Gamma_{even} (0,A_j)
\;=\; \Gamma (0, A_j) 
\ee
where the last equality proceeds from the fact that there is no
odd part for $\theta = 0$. We can then take the even contribution out of 
the integral over $\theta$, obtaining
\be
\frac{{\cal Z}_q (A)}{{\cal Z}(A)} \, \simeq \, 
\int_{-\pi}^\pi \frac{d \theta}{2 \pi} \, 
e^{- i \theta q - \Gamma_{odd} (\theta, A) } \;.
\label{ratio}
\ee
Note that in the last expression ${\cal Z} (A) \equiv 
\exp [-\Gamma (0, A_j)]$ is the partition function
in the presence of a magnetic field in the {\em canonical\/}
ensemble. This shows that the specific properties of 
the fixed charge ensemble when $\beta |M|$ is large are 
determined by $\Gamma_{odd}$. Equivalently, in terms of the 
respective free energies $F \equiv - \frac{1}{\beta} \log Z$,
\be
F_q \,-\, F \;\simeq\; - \frac{1}{\beta} 
\log \left\{ \int_{-\pi}^\pi \frac{d \theta}{2 \pi} \, 
e^{- i \theta q - \Gamma_{odd} (\theta, A) }  \right\}\;.
\label{diff}
\ee
Now we can consider the behaviour of (\ref{ratio}) for 
different limits: When $\beta M \to \infty$, as the parity
anomaly term  cancels the induced term coming from the explicit 
parity breaking mass $M$, so that $\Gamma_{odd}$ tends 
to zero. This means that, when $\beta M \to \infty$,
\be
F_q \,\simeq \, F \,\to\,- \frac{1}{\beta}\log [\delta_{q,0}] \; 
\ee
The meaning of this equation is clear, ensembles with non-zero
charge are separated by an infinite free energy barrier, and
only the zero charge one is physically possible.
When $\beta |M|$ is large but not necessarily zero, ensembles
with $q \neq 0$ are possible, and we shall discuss them now.
We first note that, due to Parseval's identity, as $\Gamma_{odd}$
is purely imaginary, we have the sum rule
\be
1 \;=\; \sum_{n=-\infty}^{n=+\infty} 
|\frac{{\cal Z}_q (A)}{{\cal Z}(A)}|^2
\ee
whose physical meaning in this case is that only a very few number
of $q's$ shall be accessible with a finite free energy. 
We shall now derive a more convenient formula for (\ref{ratio}) in 
terms of the dimensionless parameters of the theory. We define
the dimensionless quantity $b \equiv \frac{M}{|M|}\frac{e \Phi}{2 \pi}$,
which essentially measures the magnetic flux in units of the elementary 
flux quantum $\frac{e \Phi}{2 \pi}$.
We then note that after some elementary algebra, (\ref{ratio})
may be written as follows:
\be
\frac{{\cal Z}_q (A)}{{\cal Z}(A)}\;=\; \int_{-\pi}^\pi   
\frac{d \theta}{2 \pi} \; e^{- i \theta 
(q - \frac{b}{2})} \,
\left( \frac{1 +e^{-2 \beta |M|} e^{- i \theta}}{e^{-2 \beta |M|} + 
e^{- i \theta}} \right)^{\frac{b}{2}} \;.
\ee 
The change of integration variable $z = e^{- i \theta}$ maps the 
integration path to a unit circle in the complex plane:
\be
\frac{{\cal Z}_q (A)}{{\cal Z}(A)}\;=\; \frac{i}{2 \pi}
\oint_C \frac{d z}{z} \; z^{q - \frac{b}{2}} \,
\left( \frac{1 +e^{2 \beta |M|} z }{e^{2 \beta |M|} + z} 
\right)^{\frac{b}{2}}
\ee
which, if $b$ is even, say $b = 2 k$ for an integer $k$, can be evaluated 
as the sum of the residues over the two poles inside the unit circle.
The result of this procedure may be put as
$$
\frac{{\cal Z}_q (A)}{{\cal Z}(A)}\;=\;
\frac{\Theta (q \leq k)}{(k -q)!} \lim_{z -> 0}
\frac{d^{k - q}}{d z^{k - q}}
\left[ 
\frac{1 +e^{2 \beta |M|} z }{e^{2 \beta |M|} + z}
\right]^k \;+
$$
\be
\frac{\Theta (k < 0)}{(k -1)!} \lim_{z \to - e^{-2 \beta |M|}}
\frac{d^{k-1}}{d z^{k-1}} \left[ 
z^{q-k-1} (1 + e^{-2 \beta |M|}z)^{-k}
\right] \;,
\label{even}
\ee
where the symbol $\Theta (inequality)$ is defined to be one if
the inequality is true, and zero otherwise. This is not a closed form 
but may be exactly evaluated for any set of values for $q, k$ and 
$\beta M$. Note than when the sign of the magnetic flux
is the same as the one the mass, $k$ becomes positive, and so
the second term in (\ref{even}) vanishes:
\be
\left[\frac{{\cal Z}_q (A)}{{\cal Z}(A)} \right]_{k>0}\;=\;
\frac{\Theta (q \leq k)}{(k -q)!} \lim_{z -> 0}
\frac{d^{k - q}}{d z^{k - q}}
\left[ 
\frac{1 +e^{2 \beta |M|} z }{e^{2 \beta |M|} + z}
\right]^k \;.
\ee
From a numerical evaluation of this expression, we see that
finite temperature effects strongly affect  the properties
of the free energy. In particular, for $\beta |M|$ of order 
$1$, the maximum of the ratio $\frac{{\cal Z}_q (A)}{{\cal Z}
(A)}$ is reached when $q$ is equal to $k$. This means that, when
the system is heated, the Chern-Simons term makes states
with total charge proportional to the total flux more
convenient energetically. The situation is qualitatively similar
for an odd number of fluxes, though we could only check that
numerically.
We shall now discuss the issue of the meaning of the fixed-charge
ensembles in the `anomalous' case, namely, when the effective
action is not invariant under large gauge transformations. 
Invariance under large gauge transformations is, in our
case, tantamount to periodicity in $\theta$. Coming back to
the definition of the fixed-charge partition function 
(\ref{defzt}), we may say that the effect of the induced
Chern-Simons term, in the anomalous case, is equivalent to
having states of {\em fractional\/} charge.  And indeed,
a trivial way of recovering a fixed-charge ensemble for this
case also would be to fix the total charge to a fractional
value.  An equivalent way of saying this is that, if the 
parity anomaly term is lacking, the effective action is
no longer $2 \pi$-periodic , but has a period of $4 \pi$,
what can be attached to a redefinition of the charge operator. 

It is important to realize that, had we used the perturbative 
result \cite{NS}-\cite{I} for the induced Chern-Simons term, no structure 
such as the ones we are seeing here would have arise. Indeed,
the very problem of defining the fixed charge ensemble would be
ill-defined, since for the perturbative Chern-Simons term
the periodicity in ${\tilde A}_3$ is lost, and cannot be
rescued by a simple interpretation in terms of a fractional charge.
If the perturbative result is used there is no periodicity
whatsoever. 
We shall here extend the previous discussion to the non-Abelian case. 
It seems that we should now deal with a large number of fixed charges. 
However, one should remember~\cite{kapu} that in Statistical
Mechanics not all the charges can be fixed but only a subset of them
that commutes with the Hamiltonian and with all the other 
charges~\footnote{This also happens in the grand canonical ensemble,
where only such a subset of charges may carry chemical potentials.}.
Thus, when considering the partition function for fermions in a
non-Abelian magnetic background, $\theta$ will have to be a matrix
commuting with the spatial components $A_j$ of the non-Abelian
gauge field. By identifying again $\theta$ with $A_3$, this is 
precisely the kind of configuration that has been considered in 
\cite{frs2}. Obviously, the number of different integrations
will depend on the group. For example, for $SU(2)$ there will only
be one such a $\theta$, and we have the analogous of (\ref{ratio}),
the only change being a different expression for $\Gamma_{odd}$.
In the general case, ${\vec \theta} \equiv (\theta^a)$ will have 
a number $f$ of components in internal space corresponding to the 
`directions' of the fixed charges. Obviously the maximum allowed 
value for $f$ shall depend on the group, for example, $f=1$ for
the group $SU(2)$.
Denoting by ${\vec q}$ the values of such charges,
the corresponding partition function is, in the same approximation
we used for the Abelian case,
\be
\frac{{\cal Z}_{\vec q} (A)}{{\cal Z}(A)} \, \simeq \, 
\int_{-\pi}^\pi \cdots \int_{-\pi}^\pi \,
\frac{d {\vec \theta}}{(2 \pi)^f} \, 
e^{- i {\vec \theta} \cdot  {\vec q} - \Gamma_{odd} 
({\vec \theta}, A) } \;,
\ee
where
\be
\Gamma_{odd} \;=\;
\frac{ig}{4\pi} tr \left(\arctan[\tanh(\frac{\beta M}{2}) \tan(
\theta )]
\, \int d^2x \varepsilon_{ij} F_{ij} \right) \;,
\ee
where $\theta \equiv \theta^a \tau^a$, and we are using the same
conventions as in \cite{frs2}.
We conclude by saying that the use of the non-perturbative 
parity-breaking term in the effective action is 
essential for the definition of fixed-charge ensembles in
$2+1$ dimensions. Even if one is going to assume that 
the fermions are coupled to a dynamical gauge field, the
constant $\theta$ will appear coupled to the fermionic
current together with the third component of the gauge
field, and again $\theta$ cannot be assumed to be small
since periodicity (and the interpretation as a fixed-charge
ensemble) would be lost. 
\underline{Acknowledgments}: 
The author acknowledges G.~L.~Rossini and F.~A.~Schaposnik for
reading this manuscript.

\end{document}